\documentstyle[prd,aps,psfig,floats]{revtex}
\tighten

\begin{document}
\draft
\preprint{} 
\twocolumn[\hsize\textwidth\columnwidth\hsize\csname@twocolumnfalse\endcsname 
\title{Cosmic Microwave Background Polarization Signals from 
Tangled Magnetic Fields}
\author{T. R. Seshadri$^1$ and Kandaswamy Subramanian$^2$ \\
$^1$ Harish-Chandra Research Institute, Chhatnag Road, Jhusi,
Allahabad 211019, India\\
$^2$National Centre for Radio Astrophysics, TIFR,
Poona University Campus, Ganeshkhind, Pune 411 007, India. }
\maketitle

\begin{abstract}
Tangled, primordial cosmic magnetic fields create small rotational 
velocity perturbations on the last scattering surface (LSS) of 
the cosmic microwave background radiation (CMBR). For fields which 
redshift to a present value of $B_0 = 3\times 10^{-9}$ Gauss,
these vector modes are shown to generate polarization anisotropies of 
order $0.1\mu K - 4 \mu K$ on small angular scales ($ 500 < l < 2000$), 
assuming delta function or a power law spectra with $n=-1$. About $200$ 
times larger signals result for $n=2$ spectra.  Unlike inflation generated, 
scalar modes, these signals are dominated by the odd parity, B-type 
polarization, which could help in their detection.
\end{abstract}

\maketitle
\date{\today}
] \renewcommand{\thefootnote}{\arabic{footnote}} \setcounter{footnote}{0}

Magnetic fields in astronomical objects, like galaxies, could grow
by the amplification of small seed fields by turbulent dynamo action
\cite{dynam}. However, the need to produce magnetic helicity in galaxies
seems to severely constrain the efficiency of such dynamo action
\cite{dyndeb}. Alternatively, the galactic field could be 
of primordial origin \cite{primhyp}, although,
there is no compelling mechanism for producing the
required field \cite{euseed}. A primordial field that redshifted to
a present value of $\sim 10^{-9}$ Gauss, tangled
on galactic scales, could also significantly affect galaxy formation
\cite{wasser,ksjdb}. It is of considerable interest, therefore, to find
ways of constraining or detecting such fields \cite{euseed,kron}. 
Observations of anisotropies in the CMBR, provide a potentially 
powerful constraint. Indeed, the CMBR temperature 
isotropy can be used to place limits, of order several
nano-Gauss, on both the uniform \cite{barrow} and tangled components 
of the magnetic fields \cite{ksjd2,dur}.
However, temperature anisotropies in inflationary 
models of structure formation, will be dominated by
"non-magnetic", scalar modes. So the detection of a magnetic 
field induced signal is likely to be difficult, except possibly
on scales smaller than the Silk damping scale \cite{ksjd2}. 
Here we point out the advantage of using alternatively,
the polarization anisotropy.
Note that scalar perturbations only produce, 
what is known as E-type polarization 
anisotropy. However, as we show here, tangled magnetic fields 
which drive significant vector perturbations, also lead to a 
distinctive, significant and potentially detectable
B-type polarization anisotropy of the CMBR. 
This could help in separating their contribution
from scalar contributions, to detect/constrain such tangled,
primordial fields.

Polarization of the CMBR arises from the Thomson scattering of radiation
from free electrons, and is sourced by the quadrupole component of
the CMBR anisotropy. The evolution equations of temperature 
and polarization anisotropy for vector perturbations have been derived 
in detail in Ref.\cite{huwhit}, in the total angular momentum 
representation. We use their results extensively below.
The anisotropy in the temperature and polarization 
is expanded in terms of tensor spherical harmonics. This enables one to 
write evolution equations, for the moments, $\Theta_l^{(m)}$,
$E_l^{(m)}$ and $B_l^{(m)}$, of the temperature 
anisotropy ($\Delta T/T$), the electric (E-) type and the odd parity, 
magnetic (B-) type polarization anisotropies, respectively. 
Here $l$ stands for the multipole number 
and $m=0, \pm 1, \pm 2$, respectively, for scalar, vector and tensor 
perturbations. For vector perturbations ($m=\pm 1$), 
the magnetic type contribution dominates the 
polarization anisotropy \cite{huwhit}. Its evolution is given by 
(Eqn. (77), (62) and (18) of \cite{huwhit} ), 
\begin{equation}
{B_l^{(m)}(\tau_0,k) \over 2l +1} 
= - \sqrt{6} \int_0^{\tau_0}d\tau g(\tau_0,\tau )
P^{(m)} \beta_l^{(m)}(k(\tau_0 - \tau))
\label{bdef}
\end{equation}
where $P^{(m)}(k,\tau) = [\Theta_2^{(m)} - \sqrt{6} E_2^{(m)}]/10$ and 
$\beta_l^{(1)}(x) = \sqrt{(l-1)(l+2)}j_l(x)/2x$,
with $j_l(x)$ the spherical Bessel function of order $l$. 
The `visibility function', $g(\tau_0,\tau )= 
\dot \kappa(\tau) \exp [{-\int_{\tau}^{\tau_0}
\dot \kappa (\tau ^{\prime})d\tau^{\prime}}]$, determines the probability 
that a photon reaches us at the conformal time $\tau _0$ if it was 
last scattered at the epoch $\tau $. 
Here $\dot\kappa (\tau )=n_e(\tau )\sigma _Ta(\tau )$, $n_e$ the electron 
number density, $\sigma _T$ the Thomson cross section, and $a(\tau )$ 
the cosmic scale factor normalised to unity at the present. 
We assume a flat universe throughout.

For standard recombination, $g$ is sharply peaked about 
the time of recombination. We therefore need to calculate $P^{(1)}$, 
and hence the quadrupole anisotropies around this epoch of last scattering. 
These can be analytically estimated using the tight-coupling 
approximation, $k/\dot \kappa = k L_{\gamma} \ll 1$. 
Here $L_{\gamma}(\tau) = (\dot \kappa)^{-1}$ is the co-moving, photon mean
free path.  First, to leading order in this approximation, 
we have zero quadrupoles, and  a dipole $\Theta_1^{(1)} = v_B^{(1)}$, 
where $v_B^{(1)}$ is the magnitude of the (vector or rotational 
component of) baryon fluid velocity field, in Fourier space.  
However, to the next order the quadrupole is
not zero. It is generated from the dipole at the `last but one' 
scattering of the CMBR. Using the moments of the Boltzmann equations 
for the temperature and polarization anisotropies 
(Eq. (60), (63) and (64) of \cite{huwhit}), 
one gets $\Theta_2^{(1)} = -4 E_2^{(1)}/\sqrt{6} 
=  4 k L_{\gamma} v_B^{(1)}/(3\sqrt{3})$ and hence  
$ P^{(1)} = \Theta_2^{(1)}/4 = k L_{\gamma} v_B^{(1)}/(3\sqrt{3})$.
Using this in Eq. (\ref{bdef}) gives an estimate of $B_l^{(1)}$, and
the angular power spectra $C_l^{BB}$ due to B-type polarization
anisotropy. We use Eq. (56) of \cite{huwhit} to relate $C_l^{BB}$ interms
of $B_l^{(1)}$ and get
\begin{eqnarray}
C_l^{BB} &&=2{(l-1)(l+2) \over l(l+1)}
{4\pi \over 9} \int_0^\infty { k^2dk \over 2\pi^2} 
{l(l+1) \over 2} \times \quad \nonumber \\
 && <\left \vert \int_0^{\tau _0}d\tau g(\tau _0,\tau )
{ kv_B^{(1)}(k,\tau) \over\dot \kappa(\tau)}
{\frac{ j_l(k(\tau _0-\tau ))}{k(\tau _0-\tau )}}\right \vert^2>. 
\label{deldef}
\end{eqnarray}
Here we have included an extra factor of $2$, since we have 
to sum over the power in both $m=+1$ and $m=-1$ contributions.
The above expression for $C_l^{BB}$ is very closely related to 
Eq. (1) of Ref.\cite{ksjd2} (henceforth, Paper I),
for the temperature power spectrum $C_l$, due to tangled magnetic fields.
One can make the same approximations as made
there, to obtain an analytic estimate of $C_l^{BB}$. 

Firstly, it suffices to approximate the visibility 
function as a Gaussian: $g(\tau _0,\tau)=
(2\pi \sigma ^2)^{-1/2}\exp[-(\tau-\tau _{*})^2/(2\sigma ^2)]$, 
where $\tau _{*}$ is the conformal epoch of ``last scattering'' 
and $\sigma $ measures the width of the LSS. 
Using the expressions given in Ref. \cite{husug}, we estimate 
$\tau _{*}\sim 176.2h^{-1}Mpc$ and $\sigma =11.6h^{-1}Mpc$. 
($h$ is the Hubble constant in units of $100$ km s$^{-1}$ Mpc$^{-1}$.
We take $h=0.75$ henceforth and adopt a baryon density parameter 
$\Omega_b=0.02h^{-2}$.)
The dominant contributions to the integral over $\tau $ in 
Eq. (\ref{deldef}) then comes from a range $\sigma $ around the 
epoch $\tau =\tau _{*}$. Further $j_l(k(\tau _0-\tau ))$ picks 
out $(k,\tau )$ values in the integrand which have 
$k(\tau _0-\tau )\sim l.$

Let us also assume that $ kL_{\gamma}v_B^{(1)}(k,\tau)$ 
varies slowly with $\tau$ (around $\tau \sim \tau_*$), 
and slowly with $k$ around $k\sim l/R_{*}$, the regions
which contribute dominantly to the integrals in Eq. (\ref{deldef}).
(Here $R_{*}=\tau _0-\tau _{*}$).
Then following exactly the arguments detailed in paper I,
we get for $k\sigma << 1$, the analytical estimate,
\begin{equation}
{\frac{l(l+1)C_l^{BB}}{2\pi }}\approx 
\left({kL_{\gamma}(\tau_{*}) \over 3}\right)^2
{\frac \pi 4}\Delta _v^2(k,\tau _{*})|_{k=l/R_{*}}.  \label{powlar}
\end{equation}
Here, $\Delta _v^2=k^3<|v_B^{(1)}(k,\tau _{*})|^2
+ |v_B^{(-1)}(k,\tau _{*})|^2>/(2\pi ^2)$ 
is the power per unit logarithmic interval of $k$, residing in the
{\it net} rotational velocity perturbation.
And in the other limit, $k\sigma >>1$, we get 
\begin{equation}
{\frac{l(l+1)C_l^{BB}}{2\pi }}\approx 
\left({kL_{\gamma}(\tau_*) \over 3}\right)^2
{\frac{\sqrt{\pi }}4}{\frac{\Delta_v^2(k,\tau _{*})}{k\sigma
}}|_{k=l/R_{*}}.  
\label{powsmal}
\end{equation}
For small wavelengths, $C_l^{BB}$ is suppressed by a 
$1/k\sigma $ factor due to the finite thickness of the LSS. 
Note that in both cases, the polarization 
anisotropy, $\Delta T_P^{BB}(l) \approx (kL_{\gamma}(\tau_*)/ 3) 
\times \Delta T(l)$, where, $\Delta T(l)$ is the temperature anisotropy
computed in Paper I. 

To evaluate $C_l^{BB}$, one needs to estimate $v_B^{(1)}$,
for a general spectrum of magnetic inhomogeneities. 
We assume the magnetic field to be initially
a Gaussian random field. On galactic scales and above, the induced
velocity is generally so small, it does not lead to any appreciable 
distortion of the initial field \cite{ksjdb}. So, to a very good 
approximation, the magnetic field simply redshifts as, 
${\bf B}({\bf x},t)={\bf b}_0({\bf x})/a^2$. 
The Lorentz force associated with the tangled field is then 
${\bf F}_L=({\bf \nabla }\times {\bf b}_0)\times {\bf b}_0/(4\pi a^5)$,
which pushes the fluid, creating rotational velocity
perturbations. Further, the magnetic stresses, say $\Pi_B$, 
can lead to metric perturbations. We focus on scales
larger than the photon mean-free-path at decoupling, and describe the
viscous effect due to photons, in the diffusion approximation.
The Fourier transform of the linearised Euler equation for $v_B^{(1)}$,
is given by \cite{huwhit,ksjdb},
\begin{eqnarray}
\left( {\frac 43}\rho _\gamma +\rho _b\right) 
{\frac{\partial (v_B^{(1)} - V)}{\partial t}}+ 
{\frac{\rho _b}a}{\frac{da}{dt}}  (v_B^{(1)} &-& V) 
+{\frac{k^2\eta }{a^2}}v_B^{(1)}\nonumber \\
&=& {\frac{F_B^{(1)}}{4\pi a^5}}.  \label{eulerk}
\end{eqnarray}
Here $V(k,t)$ is the vector component of the metric perturbation 
($t$ is comoving proper time), $\rho_\gamma$ the photon density,
$\rho_b$ the baryon density, and $\eta =(4/15)\rho _\gamma l_\gamma $ 
the shear viscosity coefficient associated with the damping due 
to photons, whose mean-free-path is $l_\gamma = L_{\gamma} a(t) $. 
$F_B^{(1)}$ is the $m=1$ component (defined as in ref. \cite{huwhit}) of
$P_{ij}F_j$, the rotational part of the Lorentz force.
We have defined the Fourier transforms of the magnetic field as, ${\bf
b}_0(%
{\bf x})=\sum_{{\bf k}}{\bf b}({\bf k})\exp (i{\bf k}.{\bf x})$. 
Since the Lorentz force is non-linear in ${\bf b}_0({\bf x})$,
this leads to the mode-coupling term ${\bf F}({\bf k})=\sum_{{\bf p}}[%
{\bf b}({\bf k}+{\bf p}).{\bf b}^{*}({\bf p})]{\bf p}-[{\bf k}.{\bf
b}^{*}(%
{\bf p})]{\bf b}({\bf k}+{\bf p})$. The projection tensor, $P_{ij}({\bf
k}%
)=[\delta _{ij}-k_ik_j/k^2]$ projects ${\bf F}$ onto its transverse
components (perpendicular to ${\bf k}$ ).

The comoving Silk scale at recombination, 
$L_S =k_S^{-1} \sim (l_{\gamma}(t_*) t_*)^{1/2}/a(t_*)
\sim 6.8 Mpc$, separates scales for which the damping term 
in (\ref{eulerk}), is important ($ kL_S>>1$) from those for which it 
is negligible ($kL_S<<1$) \cite{clar}. 
We can solve Eq.(\ref{eulerk}) 
analytically, in these two limits. For $kL_s<1$, and when the fluid 
starts from rest ($v_B^{(1)}(\tau _i)=0$), the damping due to the photon 
viscosity can be neglected compared to the Lorentz
force. Integrating Eq.(\ref{eulerk}) gives 
$v_B^{(1)}=V + G_B^{(1)}(\tau -\tau _i)/(1+S_{*})$, where 
we have defined $G_B^{(1)}=3F_B^{(1)}/[16\pi\rho_0]$, 
with $\rho_0$ the redshifted
present day value of $\rho_\gamma$, and 
$S_{*}=(3\rho _b/4\rho _\gamma)(\tau _{*})\sim 0.73(\Omega_b/0.02h^{-2})$. 
The metric perturbation term $V$ is also smaller than the Lorentz force
driven contribution to $v_B^{(1)}$, for large $l$ by a factor
$\sim (l/24)^{-2} h^{-2}$ 
(see \cite{vector}); and so makes a negligible contribution 
to $C_l^{BB}$, for the small angular scales ($l>400$) considered here.
In the other limit, with $kL_s>>1$, we can use the terminal-velocity 
approximation, neglecting the inertial terms in the Euler equation, 
and balance the Lorentz force by friction. This gives 
$v_B^{(1)}=(G_B^{(1)}/k)(kL_\gamma /5)^{-1}$, independent of $V$.

We also need to specify the spectrum of the tangled magnetic 
field, $M(k)$. We define, 
$<b_i({\bf k})b_j({\bf q})>=\delta _{{\bf k},{\bf q}}
P_{ij}({\bf k})M(k)$, where $\delta _{{\bf k},{\bf q}}$ 
is the Kronecker delta which is non-zero only for ${\bf k}={\bf q}$. 
This gives $<{\bf b}_0^2>=2\int (dk/k)\Delta _b^2(k)$, where $\Delta
_b^2(k)=k^3M(k)/(2\pi ^2)$ is the power per logarithmic interval in $k$
space residing in magnetic tangles, and we replace the summation over 
$k$ space by an integration. The ensemble average $<|v_B^{(1)}|^2>$, and
hence the $C_l^{BB}$s, can be computed in terms of the magnetic 
spectrum $M(k)$. It is convenient to define a dimensionless spectrum, 
$h(k)\equiv \Delta _b^2(k)/(B_0^2/2)$, where $B_0$ is a fiducial 
constant magnetic field. The Alfv\'en velocity, 
$V_A$, for this fiducial field is, 
\begin{equation}
V_A ={\frac{B_0}{(16\pi\rho_0/3)^{1/2}}} \approx 
3.8\times 10^{-4}B_{-9}. 
\label{alfvel}
\end{equation}
where $B_{-9} = (B_0/10^{-9} {\rm Gauss})$.
Also, as a measure of the B-type CMBR polarization anisotropy 
induced by the tangled magnetic field, we define the quantity 
$\Delta T_P^{BB}(l)\equiv [l(l+1)C_l^{BB}/2\pi]^{1/2}T_0$. 

Note that when $k\sigma<1$ (or $l < 500$), 
one also generally has
$kL_S <1$ \cite{clar}. The resulting $\Delta T_P^{BB}$ can be
estimated using Eq.(\ref{powlar}), and ignoring viscous damping. 
A lengthy calculation gives, for such scales,
$\Delta T_P^{BB}(l)=T_0 (\pi/32)^{1/2} I(k) 
[k^2V_A^2\tau_*L_{\gamma}(\tau_*)/3(1+S_{*})]$, where, $l=kR_{*}$.
For scales with $kL_S>1$ ($l > 1150$), we can use Eq.(\ref{powsmal}),
and $v_B^{(1)}=(G_B/k)(kL_\gamma /5)^{-1}$. 
A similar calculation to that above gives, 
$\Delta T_P^{BB}(l)=T_0 (5\pi^{1/4}/12\sqrt{2})
(I(k) V_A^2/(k\sigma)^{1/2})$.
The function $I^2(k)$ is a
dimensionless mode-coupling integral given by \cite{ksjd2} 
\begin{eqnarray}
I^2(k) &=&\int_0^\infty {\frac{dq}q}\int_{-1}^1d\mu {\frac{h(q)h(|({\bf
k}+%
{\bf q})|)k^3}{(k^2+q^2+2kq\mu )^{3/2}}}  \nonumber \\
&&\ \times (1-\mu ^2)\left[ 1+{\frac{(k+2q\mu )(k+q\mu )}{(k^2+q^2+2kq\mu
)}}%
\right]   \label{modint}
\end{eqnarray}
where $|({\bf k}+{\bf q)}|=(k^2+q^2+2kq\mu )^{1/2}$. 
Putting in numerical values we estimate for 
$l<500$ and $l >1150$ respectively,
\begin{eqnarray}
\Delta T_P^{BB}(l) &\approx & 0.21\mu K \ I({\frac l{R_{*}}})
\left( {\frac{B_{-9}}3}\right) ^2\left( {\frac l{400}}%
\right)^2  \nonumber \\
&\approx &0.93\mu K \ I({\frac l{R_{*}}})
\left( {\frac{B_{-9}}3}\right) ^2\left( {\frac l{1500}}%
\right) ^{-1/2} 
\label{largT}
\end{eqnarray}
Further, for intermediate scales, $500 < l < 1150$, an estimate
using Eq.(\ref{powsmal}), but ignoring viscous damping, gives
$\Delta T_P^{BB}(l) \sim 0.5 \mu K (l/800)^{3/2} I(l/R_*)(B_{-9}/3)^2$.

If the magnetic spectrum has a single scale, 
with $h(k)=k\delta _D(k-k_0)$, where $\delta _D(x)$ is the Dirac delta 
function, $<{\bf b}_0^2>=B_0^2$ and the mode-coupling integral can be 
evaluated exactly. We find $I(k)=(k/k_0)[1-(k/2k_0)^2]^{1/2}$, 
for $k<2k_0$, and  $I(k)=0$ for larger $k$. So $I(k) \sim 1$ 
for $k\sim k_0$, with a maximum $I(\sqrt{2}k_0)= 1$. 
For $B_{-9} \sim 3$, one then expects a RMS
$\Delta T_P^{BB}\sim 0.2\mu K - 1 \mu K$, 
depending on  $k_0$ and $l$. 
We can also consider power law spectra, $M(k)= A k^n$
cut-off at $k=k_c$, where $k_c$ is the Alfven-wave
damping length-scale \cite{ksjdb,alfdamp}.
We fix $A$ by demanding that the field 
smoothed over a "galactic" scale $k_{G}=1h {\rm Mpc}^{-1}$, 
(using a sharp $k$-space filter) is $B_0$, giving
$h(k) = (n+3)(k/k_G)^{3+n}$ ($n > -3$).
We then find for $k<<k_c$ 
(as is relevant for $l < 2000$), and $n > -3/2$, 
$I^2(k) = (28/15)((n+3)^2/(3+2n))(k/k_G)^3(k_c/k_G)^{3+2n}$.
Using this in Eq.(\ref{largT}), we find 
$\Delta T_P^{BB} \sim 0.16 \mu K (l/400)^{7/2}$ for $l < 500$ and
$\Delta T_P^{BB} \sim 5 \mu K (l/1500)$ for $l > 1150$, for $n=-1$
and $B_{-9} \sim 3$. Much larger signals result
for larger $n=2$, say, and same $B_0$, with 
$\Delta T_P^{BB} \sim 9.6 \mu K (l/400)^{7/2}$ for $l < 500$ and
$\Delta T_P^{BB} \sim 308.5 \mu K (l/1500)$ for $l > 1150$.

\begin{figure}
\begin{picture}(200,145)
\psfig{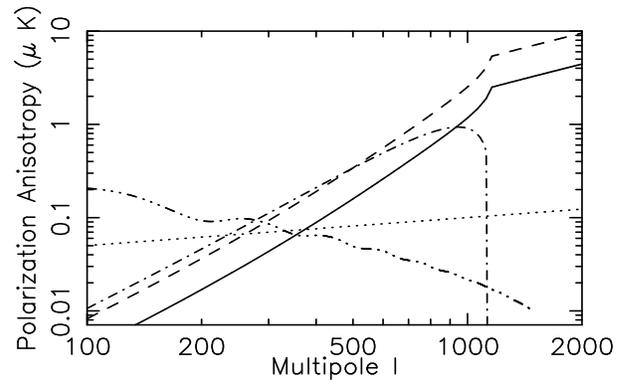}
\end{picture}
\caption{ $\Delta T_P^{BB}$ versus $l$ plot for different model $P(k)$
taking $B_{-9}=3$. The dashed-dotted line is for the delta 
function spectrum with $k_0 =0.075$ Mpc$^{-1}$. The solid curve, is for the 
power law spectrum, with $n=-1$, while the 
dashed curve, gives $\Delta T_P^{BB}/100$ for $n=2$.
For comparison, we also show (dash-triple dotted line),
the tensor contribution to B-type polarization, 
in a standard cold dark matter model, computed with  
the default parameters of CMBFAST, except changing the scalar
spectral index to $n_S=0.9$. The dotted line, gives the contribution 
of galactic foreground emission.}
\end{figure}

To complement the analytics, we have also computed $\Delta T_P^{BB}$ for 
the above spectra, by evaluating the $\tau$ and $k$ integrals in 
Eq.(\ref{deldef}) numerically (but using analytical approximations
to $v_B^{(1)}$). The results are shown in Figure 1.
We see that for $B_0\sim 3\times 10^{-9}G$, one expects a RMS
B-type CMBR polarization anisotropy of order $0.1\mu K - 4 \mu K$ 
for $ 500 < l < 2000$, for delta function or a power law
spectra with $n=-1$, and $\sim 200$ times larger signals
for $n=2$ spectra.  These values compare reasonably with
the analytical estimate from Eq.(\ref{largT}) (see also \cite{ldep}).
Further, both the tensor contribution of 
typical inflationary models ( from CMBFAST \cite{sel}), 
and the foreground contribution (dominated by galactic dust polarized 
emission for high frequencies cf.\cite{shiv}) are subdominant to 
the magnetic field induced signal, for large $l > 400$.
Clearly with the sub-micro Kelvin sensitivities expected from Planck
\cite{plank}, these signals can be detected.

Earlier work also emphasised the possibility of Faraday rotation
and the depolarization of the CMBR due to differential Faraday rotation
in a tangled magnetic field \cite{harhayzal}. Note that the average 
Faraday rotation (in radians) between Thomson scatterings is given by 
$F = 3B_0/(2\pi e\nu_0^2) \approx 0.23 
(B_0/3 \times 10^{-9} G)(\nu_0/30 GHz)^{-2}$, where
$\nu_0$ is the observed frequency.
The CMB could then become
significantly de-polarised due to this effect, for 
$\nu_0 < 16.4 GHz (B_0/3 \times 10^{-9} G)^{1/2}$,
but Faraday rotation effects will be negligible for say $\nu_0 > 40 GHz$, or 
the higher frequency instruments of the Planck Satellite.

Note that scalar modes can also be induced by tangled 
magnetic fields and generate a purely E-type polarization.
These are however of smaller amplitude than vector modes, 
due to the larger restoring force contributed by the radiation-baryon 
fluid pressure \cite{ksjdb}. They are also strongly damped on scales 
smaller than the Silk scale, $L_S$, while the vector mode, 
survives on much smaller scales $> V_AL_S$ \cite{ksjdb}, 
or larger $l$ (see Figure 1).
The tensor mode can also contribute to $\Delta T_P^{BB}$,
but one can show that their effect is only important at
large angular scales. Further, $\Delta T_P^{BB}$ could also be
generated, after the universe gets re-ionized, though
from the recent detection of multiple doppler peaks, the optical
depth for scattering at re-ionization is probably very small \cite{max}.
A detailed computation of these effects will be presented elsewhere.

In conclusion, we have identified here, a new physical
effect of tangled magnetic fields; that they can produce distinctive
and potentially detectable B-type polarization anisotropy on arc 
minute scales. From Eq.(\ref{largT}) and Figure 1, we see that a tangled
field with $B_0\sim 3\times 10^{-9}G$, induces a RMS
B-type CMBR polarization anisotropy of order $0.1 \mu K - 900 \mu K$ or
larger, depending on $M(k)$ and $l$.
The anisotropy in hot/cold spots could be several times
larger, because the non-linear dependence of $C_l^{BB}$ on $M(k)$ will
imply a non-Gaussian statistics for the anisotropies (see Paper I). 
Further in standard models all the $C_ls$ have a sharp cut-off for $%
l>R_{*}/L_S$, due to Silk damping but strong damping of Alfv\'enic
modes is expected only on scales smaller than $V_AL_S$
\cite{ksjdb}. Finally, since tangled magnetic fields produce 
predominantly B-type polarization, which are also dominant at large $l$,
they can be distinguished from those produced by inflationary
scalar and tensor perturbations.
Satellite bourne experiments like Planck \cite{plank},
with the sub-micro Kelvin sensitivities
should be able to detect and isolate the effects of magnetic 
fields, using CMBR polarization, if such fields indeed play a role 
in structure formation.

As this letter was prepared for (re)submission, preprints \cite{tina}
appeared which have some overlap with our work. We thank H. M. Antia
and S. Sethi for help.


\begin{references}
\bibitem{dynam}  A. A. Ruzmaikin, A. Shukurov and D. Sokoloff, {\it
Magnetic Fields of Galaxies}, Kluwer, Dordrecht (1988);
R. Beck, A. Brandenburg, D. Moss, A.  Shukurov and D. Sokoloff, 
Ann. Rev. A. \& A., {\bf 34}, 155 (1996).

\bibitem{dyndeb} A. Brandenburg, ApJ, {\bf 550}, 824 (2001),
F. Cattaneo and S. I. Vainshtein, Ap.J., {\bf 376}, L21 (1991); 
E. G. Blackman and G. F. Field, ApJ, {\bf 534}, 984 (2000); 
A. Brandenburg and K. Subramanian, A \& A, {\bf 361}, L33 (2000)

\bibitem{primhyp}  R. M. Kulsrud, IAU Symp. 140: {\it Galactic and
Extragalactic Magnetic Fields}, Reidel, Dordrecht, (1990), p527.

\bibitem{euseed}  see D. Grasso and H. R. Rubinstein, Phys. Rept. {\bf 348},
161 (2001) and references therein.

\bibitem{wasser}  I. Wasserman, Ap.J., {\bf 224}, 337 (1978)
\bibitem{ksjdb}  K. Jedamzik, V. Katalinic, and A. Olinto, Phys. Rev.
{\bf D57}, 3264 (1998). K. Subramanian and J. D. Barrow, Phys.Rev. 
{\bf D58} 083502 (1998).
\bibitem{kron}  P. P. Kronberg, Rep. Prog. Phys., {\bf 57}, 325 (1994);
\bibitem{barrow}  J. D. Barrow, P. G. Ferreira and J. Silk, Phys. Rev.
Lett. {\bf 78}, 3610 (1997).
\bibitem{ksjd2}  K. Subramanian and J. D. Barrow, Phys.
Rev Lett. {\bf 81}, 3575 (1998) (Paper I).
\bibitem{dur} R. Durrer, T. Kahniashvili and A. Yates, Phys. Rev. 
{\bf D58}, 123004 (1998); R. Durrer, P. G. Ferreira and T. Kahniashvili,
Phys. Rev. {\bf D61}, 043001 (2000). 
\bibitem{huwhit}  W. Hu and M. White, Phys.Rev. {\bf D56}, 596 (1997).
\bibitem{husug}  W. Hu, and N. Sugiyama, Ap.J., {\bf 444}, 489 (1995).
\bibitem{clar} We use \cite{huwhit2} to estimate $L_S$, adopting
$h = 3/4$ and $\Omega_b = 0.02h^{-2}$.
Although $L_S$ is of the same order as $\sigma$, their
physical origins are different. $L_S$
is the diffusion scale, governed by the
geometric mean of $\l_{\gamma}$ and the Hubble radius
at LSS, whereas $\sigma$ is sensitive to just $l_{\gamma}$. 
\bibitem{huwhit2} W. Hu and M. White, ApJ, {\bf 479}, 568 (1997).
\bibitem{vector} 
Note that $V$ satisfies the vector Einstien equation,
$\dot V + 2 (\dot a/a)V =- 8\pi G a^2\Pi_B^{(1)}/k$, where
'dot' denotes a conformal derivative (see Eq (70) of Ref. (\cite{huwhit})).
and the vector part of the stress, $\Pi_B^{(1)} = F_B^{(1)}/k$.
This integrates to give $V = -8\pi G F_B^{(1)} (\tau -\tau_i)/(k^2 a^2) $.
The ratio of $V$ to the Lorentz force term at last scattering
is then $-8\pi G (4\rho_0/3)(1+S_*)/(k^2 a(\tau_*)^2)\sim 
(l/24)^{-2} h^{-2}$. So we can neglect the $V$ term 
contribution for large $l > 100$.
\bibitem{alfdamp}
Alfv\'en modes survive damping on scales 
larger than $k_c^{-1} \sim V_A^{eff} L_S$ \cite{ksjdb}.
For computing the effective Alfven velocity, $V_A^{eff}$, 
we assume that the field smoothed over scales
$2 k_c^{-1}$ acts as an effective large-scale
field for the cut-off scale perturbations.
For $B_{-9}=3$, we get $k_c=14 Mpc^{-1}$ for $n=-1$, 
and $k_c = 5.4 Mpc^{-1}$ for $n=2$.
\bibitem{sel} U. Seljak and M. Zaldarriaga, ApJ, {\bf 469}, 437 (1996).
\bibitem{ldep} The numerically computed $l$ dependence, also agrees
reasonably with the analytical prediction for $l > 500$, but differs
somewhat at lower $l$; $ \Delta T_P^{BB} \propto l^{2.5}$,
compared to the expected $l^{3.5}$ dependence for $l < 500$. This is 
probably because the numerical integration treats more accurately
the more rapid $k$ variation of $P^{(1)}$, for small $k$,
and the effects of the finite thickeness of the LSS.
\bibitem{shiv} S. Prunet, S. K. Sethi, F. R. Bouchet and 
M. A. Miville-Deschenes, 1998, A \& A, {\bf 339}, 187. 
\bibitem{plank} cf. http://astro.estec.esa.nl/SA-general/Projects/Planck;
W. Hu, astro-ph/0002520.
\bibitem{harhayzal} A. Kosowsky and A. Loeb, ApJ., {\bf 469}, 1 (1996);
E. S. Scannapieco and P. G. Ferreira, Phys. Rev. {\bf D56}, R7493 (1997);
D. Harari, J. Hayward, M. Zaldarriaga, Phys.Rev. {\bf D55} (1997) 1841
M. Giovannini, Phys. Rev., {\bf D56}, 3198 (1997).
\bibitem{max} X. Wang, M. Tegmark and M. Zaldarriaga, astro-ph/0105091 (2001)
and references therein.
\bibitem{tina} T. Kahniashvili, A. Kosowsky, A. Mack and R. Durrer,
astro-ph/0011095 (2000); A. Mack, T. Kashniashvili, A. Kosowsky,
astro-ph/0105504 (2001).
\end{references}
\end{document}